\documentclass{ws-ijmpd}
\usepackage{epstopdf}
\usepackage{rotating}

\begin{document}

\title{Constraining values of bag constant for strange star candidates}

\author{Abdul Aziz}
\address{Department of Physics, Indian Institute of Engineering Science and Technology, Shibpur, Howrah 711103, West Bengal, India\\ aziz.rs2016@physics.iiests.ac.inx}

\author{Saibal Ray}
\address{Department of Physics, Government College of Engineering and Ceramic Technology, Kolkata 700010, West Bengal, India \& Department of Natural Sciences, Maulana Abul Kalam Azad University of Technology, Haringhata 741249, West Bengal, India\\
saibal@associates.iucaa.in}

\author{Farook Rahaman}
\address{Department of Mathematics, Jadavpur University, Kolkata
700032, West Bengal, India\\ rahaman@associates.iucaa.in}

\author{M. Khlopov}
\address{APC Laboratory, 10 rue Alice Domon et Lonie Duquet 75205, Paris, Cedex 13, France \& Center for Cosmopartcile Physics “Cosmion” and National Research Nuclear University “MEPHI” (Moscow State Engineering Physics Institute), Kashirskoe Shosse 31, Moscow 115409, Russia\\ khlopov@apc.in2p3.fr}

\author{B.K. Guha}
\address{Department of Physics, Indian Institute of Engineering Science and Technology, B. Garden, Howrah 711103, West Bengal, India\\
bkguhaphys@gmail.com}

\maketitle	
\begin{history}
\received{Day Month Year} \revised{Day Month Year}
\end{history}

\begin{abstract}
We provide a strange star model under the framework of general relativity by using a general linear equation of state (EOS). The solution set thus obtained is employed on altogether 20 compact star candidates to constraint values of MIT bag model. No specific value of the bag constant ($B$) a-priori is assumed rather possible range of values for bag constant is determined from observational data of the said set of compact stars. To do so the Tolman-Oppenheimer-Volkoff (TOV) equation is solved by homotopy perturbation method (HPM) and hence we get a mass function for the stellar system. The solution to the Einstein field equations represents a non-singular, causal and stable stellar structure which can be related to strange stars. Eventually we get an interesting result on the range of the bag constant as 41.58~MeV~fm$^{-3}< B <$319.31~MeV~fm$^{-3}$. We have found the maximum surface redshift $Z^{max}_{s}=0.63$ and shown that the central redshift ($Z_c$) can not have value larger than $2k$, where $k=2.010789 \pm 0.073203$. Also we provide a possible value of bag constant for neutron star (NS) with quark core using hadronic as well as quark EOS.  
\end{abstract}


\section{Introduction}
We have entered in the new era of astrophysics after the first successful detection of gravitational waves from a binary black hole merger~\cite{Abbott2016}. Also from a binary neutron star (NS) merger, Gamma ray burst (GRB 170817A) and Gravitational wave (GW 170817) were detected with the help of LIGO and Virgo detectors~\cite{Abbott2017a}. The detection is of greater importance in the field of astrophysics and cosmology~\cite{Abbott2017b}.  The recent claim of production of gravitational wave echo from GW 170817 can be related to strange stars and it is shown that strange star can emit gravitational wave echo~\cite{Mannarelli2018}. So, it is possible to get information about the equation of state (EOS) of quark matter by observation and analysis of gravitational waves emitted from strange star~\cite{Sotani2004}.  Again the successful detection of $f$ and lowest $\omega_{II}$ mode can constrain the values of bag constant and strange quark mass. The dark matter can be related to the presence of stable nugget of strange matter which implies keen indications in cosmology~\cite{Bhattacharyya1999,VanDevender2017}. Therefore it is very interesting to study the properties of the strange star  made of strange matter (assumed to be absolute ground state of matter)~\cite{Bodmer1971,Witten1984} which contains up, down and strange quarks. 

MIT bag model~\cite{Farhi1984} is first developed to describe the strange matter inside the strange stars. In this model, color confinement is perceived to be a bag. To estimate the size of the bag, a bag constant ($B$) is introduced. The space with lowest energy possible in it  is called as vacuum. This will be true vacuum when it has global minimum of energy and stable configuration. On the other hand, the space with local minimum of energy and unstable configurations under perturbation is called as a false vacuum. We can think of a spherical surface of certain radius and thickness as separator of interior true vacuum and exterior false vacuum~\cite{Molin2008}. So, the bag constant can be interpretated as the difference in energy density of false vacuum and true vacuum. If $B$ is added to the kinetic energy of quarks we shall get the energy density~\cite{Bordbar2012}. Again, increase in the value of $B$ decreases the quark pressure. Eventually the value of $B$ influences the structure of strange star. It is to be noted that in the MIT bag model at first $B \approx$ 55 MeV~fm$^{-3}$ is taken as the standard value which can explain the cause of spectrum of heavy mesons and light hadrons~\cite{Grand1975,Haxton1980}. The Bag constant depends on the mass of strange quarks. For massless quarks the restricted range for $B$ is 58.9-91.5 MeV fm$^{-3}$~\cite{Stergioulas2003} whereas for $m_s$=150~MeV the range for $B$ is 56-78 MeV fm$^{-3}$~\cite{Farhi1984}.  

It is seen that $B$ can have even large values. In calculations of lattice quantum chromodynamics~\cite{Satz1982} the value of B is $\approx$ 210 MeV fm$^{-3}$. Dey et al \cite{Dey1998} using phenomenological density scalar potential and quark vector interaction derive a new EOS for strang star. They compare their results to the MIT bag model EOS and showed that to describe $Her~X-1$ and $4U~1820-30$ as strange stars one need to use large value of bag constant $B \approx$ 110 MeV fm$^{-3}$. The physical validity of different relativistic EOS based model of compact stars can be checked by comparing mass-radii range obtained in the bandwidth of all resonant mass detectors. The quark stars described by the MIT bag model with $B~=$~170~MeV fm$^{-3}$ are the best fit for the mass-radii range obtained from the band of frequency of gravitational wave modes which include the bandwidth (2.8-3.4~kHz) of spherical resonant mass detectors like MiniGrail and Schenberg~\cite{Lenzi2009}. 

Burgio et al.~\cite{Burgio2002b} uses the value of $B$ which are constrained by the experimental results at CERN on quark-gluon plasma~\cite{Heinz2000,Heinz2001} which is confirmed by RHIC results~\cite{Blaizot2002}. They have showed that it is necessary to use  bag constant as density dependant parameter. Density dependent bag constant with Gaussian-like and Woods-Saxon-like parametrization is taken which gives a whole set NS configurations with possible cases for quark-hadron phase transitions and NS maximum mass to be in $1.44~ M_\odot \leq M_{max}\leq 1.7~M_\odot$~\cite{Burgio2002}. The value $B~=$~90~MeV fm$^{-3}$ is used to describe the structure of a hot NS containing a quark core~\cite{Yazdizadeh2011}. So, we may take $B$ as an effective free parameter.

The maximum masses of hybrid stars containing both hadron and quark phase are always less than $1.7~M_\odot$~\cite{Maieron2004}. Rodrigues et al.~\cite{Rodrigues2011} identified the EOS of quark star to be stiff in the presence of color superconducting quark phase. In their study it has been shown that massive compact stars of mass in the range $2~ M_\odot < M < 2.73~ M_\odot$ can be interpreted as strange star and comparatively less massive compact stars can be treated as hybrid stars. The occurrence of large masses and radii of stars possibly be explained by new and quite stiff quark matter EOS which is originated mainly due to the pairing interaction of quarks at high densities in a color superconducting phase~\cite{Weber2007,Malheiro2007,Malheiro2006}. For $B=$200~MeV fm$^{-3}$ and $B=$122~MeV fm$^{-3}$, the maximum mass of NS can reduced to be $\approx$ 2~$M_\odot$ and $\approx 1.9~M_\odot$ respectively~\cite{Akmal998}.  

In accordance with the data of CERN-SPS and RHIC, bag constant allowed to have a wide range of values~\cite{Burgio2002,Rahaman2012,Kalam2013}.  Kohri et al. assumed $RX~J185635-3754$ to be quark star and the order of upper limit of mass is determined to be $0.5-1~M_\odot$ for different values of bag parameter~\cite{Kohri2003a,Kohri2003b}. Xu et al.~\cite{Xu2003} proposed that the star $LMXB~EXO~0748-676$ may satisfy properties of strange star describing the star with MIT bag model using bag constants 60 and 110 MeV fm$^{-3}$. Deb et al.~\cite{Deb2017,Deb2018} provided a strange star model  considering MIT bag model and specific values of bag constants 83~MeV fm$^{-3}$, 100~MeV fm$^{-3}$ and 120~MeV fm$^{-3}$. 
 
According to Bordbar et al.~\cite{Bordbar2012} the bag constant plays an influential role to determine the structure and properties of strange star. It is observed that variation in its value affects the pressure, density and maximum mass of the strange star. Though $B$ has wide range of values starting from small to large but it still does not bear a definite range of values. Therefore it is essential to look for the allowed range of $B$ for the causal and stable structure of strange stars. With this motivation we have studied 20 compact stars as strange stars using MIT bag model. In the present paper we take quark EOS in general linear form and then we utilized MIT bag model as a special case. We have not used any specific value of bag constants rather we determine all the possible values of bag constants by studying compact stars. The purpose of the paper is not only to provide possible range of $B$ but also to present a general model for strange star with general linear quark EOS. We have solved TOV equation by a simple technique, known as homotopy perturbation method (HPM)~\cite{He1997,He1999,He2000,He2004,He2005,He2006,He2010}, which has many applications in astrophysics and cosmology~\cite{Shchigolev2013,Shchigolev2015a,Shchigolev2015b,Shchigolev2016,Shchigolev2017}.
  
We outline our work in the paper as follows: In Sec. 2 we express TOV equation  for the spherically symmetric spacetime as a non-linear differential equation of the mass function by using a general linear quark EOS and the Einstein field equations. The differential equation is solved by HPM and we derive an approximate analytic mass function in Sec. 3. All the features of strange star are presented, such as the density function, pressure, maximum mass, redshift, compactness, time-time component of metric etc. in Sec. 4. Then we exploited the MIT bag model as a special case of our model and find upper bound for bag constant for strange star of radius $R$ in Sec. 5. To check the physical validity of the presented model, altogether 20 compact star candidates are studied in Sec. 6. Also possible values of bag constant for NS with quark core is calculated assuming a specific value of transition density in Sec. 7. At the end, we discuss some important results of the model and make a few concluding remarks in Sec. 8.

\section{Background Mathematical Formulations}  
Metric for the static spherically system is considered in the following form
\begin{equation}
ds^{2}=-g_{tt}(r)dt^{2}+\left(1-\frac{2m(r)}{r}\right)^{-1}dr^{2}+r^{2}(d\theta^{2}+sin^{2}\theta d\phi^{2}), \label{eq1}
\end{equation}
where $g_{tt}$ is the unknown time-time component of the metric tensor. The energy momentum tensor for the isotropic fluid distribution is given by 
\begin{equation}
T_\nu^\mu= (\rho + p) u^{\mu}u_{\nu} + p g^{\mu}_{\nu}, \label{eq2}
\end{equation}
with $ u^{\mu}u_{\mu} = 1$ where $u^{\mu}$ is four velocity of fluid.

The EOS derived by Dey et al.~\cite{Dey1998} can be approximated by linear EOS provided in the MIT bag model for strange star. In this linear approximation, the EOS parameter ($\omega$) have the typical values of 0.45-0.46~\cite{Rosinska2000}. Again for MIT bag model for strange quark matter, $\omega$ is equal to 1/3 for massless strange quark and $\omega=0.289$ for $m_s$=250~MeV. In this connection it is to note that there are different EOS, such as for dust  $p=0$, for radiation of gas with the ultra relativistic particles $p=\rho/3$, for isotropic fluid composed of hadronic matter $p=\omega \rho$ and for isotropic fluid composed of quark matter $p=(\rho - 4B)/3$. We would like to study the stars composed of strange matter and for this purpose our motivations are the quark EOS in a general way so that (i) it can ensure modelling of strange star, (ii) uilization of MIT bag model through the EOS $p=(\rho - 4B)/3$ is possible, and (iii) the EOS parameter $\omega$ can have different values for different strange quark mass. 

Under the above demands, the EOS of fluid can be given by the linear relationship between the pressure and density in the following general form
\begin{equation}
p = \omega \left(\rho+\frac{b}{4\pi}\right), \label{eq3}
\end{equation}
where $b<0$ as $b=-4\pi \rho(R)$. This type of linear EOS can be used for modelling of strange star~\cite{Rosinska2000}, e.g. Sharma et al.~\cite{Sharma2002} provide relativistic model for $SAX~J1808.4-3658$ using the Vaidya-Tikekar spacetime. If we consider the MIT bag model, $b$ is related to the bag constant as $B=-b/16\pi$. Note that the appearance of $4\pi$ in Eq. 3 is for mathematical simplicity only. The constant $b$ indicates the non-zero value of the physical quantity, i.e. the energy density of a compact star at surface. 

We get the Einstein field equations using the metric (\ref{eq1}) and energy-momentum tensor (\ref{eq2}) in the following form of differential equations
\begin{equation}
\frac{2m^{\prime}}{r^{2}}=8\pi\rho, \label{eq4}
\end{equation}

\begin{equation}
\frac{2m}{r^{3}} -
\left(1-\frac{2m}{r}\right)\frac{g_{tt}^{\prime}}{g_{tt}}\frac{1}{r}
= -  8\pi p, \label{eq5}
\end{equation}

\begin{eqnarray}
-\left(1-\frac{2m}{r}\right)\left[\frac{1}{2}\frac{g_{tt}^{\prime\prime}}{g_{tt}}
-\frac{1}{4}\left(\frac{g_{tt}^{\prime}}{g_{tt}}\right)^{2} +
\frac{1}{2r}\frac{g_{tt}^{\prime}}{g_{tt}}\right] -\left (\frac{m}{r^{2}}-\frac{m^{\prime}}{r}\right)
\left(\frac{1}{r} +
\frac{1}{2}\frac{g_{tt}^{\prime}}{g_{tt}}\right) = - 8\pi p,\label{eq6}
\end{eqnarray}
where we take $G=c=1$. 

In special theory of relativity the conservation laws of energy and momentum are put in a single principle as {\it energy-momentum} in flat spacetime, i.e., $\nabla_\mu T^{\mu\nu} = 0$. However, in general relativity difficulty arises in deriving law of conservation of energy-momentum in curved spacetime~\cite{Bondi1990,Mensky2004,Wu2016}. The No-Go theorems ruled out the possibility of defining total energy over the space-like hypersurface which weakened finding the law of conservation of energy and momentum~\cite{Wu2007}. Due to unavailability of adequate expression for conservation of energy-momentum in curved spacetime for many decades, many scientists consider $\nabla_\mu T^{\mu\nu} = 0$ to be conservation of energy-momentum which works well approximately in curved spacetime~\cite{Wu2007}. In the present paper, therefore $\nabla_\mu T^{\mu\nu} = 0$ can be put in the simplified form
\begin{equation}
p^{\prime}= - \frac{(\rho + p )g_{tt}^{\prime}}{2 g_{tt}}. \label{eq7}
\end{equation}

Invoking Eq. (\ref{eq5}) in the Eq. (\ref{eq7}) we will get the Tolman-Oppenheimer-Volkoff (TOV) equation \cite{Tolman1939,Oppenheimer1939}
\begin{equation}
p^{\prime}= - \frac{(\rho + p)(m + 4\pi r^{3}p)}{r(r-2m)}. \label{eq8}
\end{equation}

Now, let us consider the line element of static and spherically symmetric system as    
\begin{equation}
ds^2= -e^{\nu}dt^2+ e^{\lambda}dr^2+r^2(d\theta^2+sin^2\theta d\phi^2),
\end{equation}
under which for an isotropic stellar system the TOV equation, i.e. Eq. (8) can be expressed in the following manner
\begin{equation}
-\frac{M_{G} (\rho +p_{r} )}{r^{2} } e^{\frac{\lambda -\nu }{2} }-\frac{dp}{dr} =0, \label{eq9}
\end{equation}
where the Tolman-Whittakar mass is given by $M_{G} (r)=\frac{1}{2} r^{2} e^{\frac{\nu -\lambda }{2} } \nu'= \frac{r^2g^{\prime}_{tt}}{2\sqrt{g_{tt}}}\sqrt{\left(1-\frac{2m(r)}{r}\right)}$. The system is in hydrostatic equilibrium as the two force terms in Eq. (\ref{eq9}), namely the gravitational force and hydrostatic force, balanced each other.

We derive a non-linear differential equation for the mass function substituting  Eq. (\ref{eq3}) and  Eq. (\ref{eq4}) in  Eq. (\ref{eq8})	
\begin{eqnarray}
m^{\prime} - \frac{1}{2}m^{\prime\prime}r-\frac{\omega b^2}{2}r^4 + m^{\prime\prime}m  -\frac{(5\omega+1)}{2\omega} \frac{mm^{\prime}}{r}\nonumber 
\\- \frac{(\omega+1)}{2}(m^{\prime})^{2} -\left(\omega+\frac{1}{2}\right)bm^{\prime}r^{2}- \frac{b}{2}mr  = 0.\label{eq10}
\end{eqnarray}

To study different features of compact star we shall find the mass function by solving the above equation with the help of a simple technique,  known as the homotopy perturbation method.

\section{Mass Function using Homotopy Perturbation Method}
It is argued by Aziz et al.~\cite{Aziz2016} that construction of a suitable homotopy structure is very important for the solution of non-linear equation. Therefore, here our only motivation is to get a suitable expression for mass function. The physical acceptance and viability of the mass function shall be discussed in later subsection 4.1 where we have utilized linear EOS and expressed arbitrary constants of integration in terms of physical parameters.

Now, we construct here a suitable homotopic relation~\cite{He2000} as 
\begin{flushleft}
\begin{eqnarray}
 m^{\prime} - \frac{1}{2}m^{\prime\prime}r-b_{1}r^4 + \epsilon\left[m^{\prime\prime}m  -\omega_{1} \frac{mm^{\prime}}{r}-\omega_{2}(m^{\prime})^{2} -b_{2}m^{\prime}r^{2}- \frac{b}{2}mr\right]=0, \label{eq14}
\end{eqnarray}
\end{flushleft}
with
\begin{eqnarray}
\left(\frac{5}{2} + \frac{1}{2\omega}\right) =\omega_{1},\nonumber 
\left(\frac{1}{2} +\frac{\omega}{2}\right) = \omega_{2}, \nonumber
\frac{\omega b^2}{2}=b_{1},\nonumber
\left(\omega+\frac{1}{2}\right)b=b_{2}.\nonumber
 \end{eqnarray}

The mass solution is assumed to be as
\begin{equation}
m = m_{0}+\epsilon m_{1}+\epsilon^2m_{2}+...~\label{eq15}
\end{equation}

After substitution of Eq. (\ref{eq15}) into Eq. (\ref{eq14}) and  making the coefficients of $\epsilon$ are equal to zero we get
\begin{equation}
\epsilon^{0}:~~~ m_{0}^{\prime} - \frac{1}{2}m_{0}^{\prime\prime}r - b_{1}r^{4}=0, \label{eq16}
\end{equation}

\begin{equation}
\epsilon^{1}:~~~ m_{1}^{\prime} - \frac{1}{2}m_{1}^{\prime\prime}r + m_{0}^{\prime\prime}m_{0} -\omega_{1} \frac{m_{0}m_{0}^{\prime}}{r} -
\omega_{2}(m_{0}^{\prime})^{2}-b_{2}m_{0}^{\prime}r^{2}-\frac{b}{2}m_{0}r =0, \label{eq17} 
\end{equation}
and so on. 

We solve Eqs. (\ref{eq16}) and (\ref{eq17}) to get 
\begin{equation}
m_{0}=-\frac{1}{5}b_{1}r^{5} + \frac{1}{3} C_{1} r^{3} + C_{2} ,\label{eq18}
\end{equation}

\begin{equation}
m_{1}= \frac{1}{9}\omega_{3}r^{9}+ \frac{1}{7}\omega_{4}r^{7}+\frac{1}{5}\omega_{5}r^{5}+\frac{1}{3}C_{3}r^{3}+ C_{4}, \label{eq19}
\end{equation}
with
\begin{eqnarray}
-\frac{b^{4}\omega^{2}}{60} \left(1+\frac{5}{2}\omega+\frac{1}{2\omega}\right)=\omega_{3},\nonumber
\frac{b^{2}\omega}{60} \left[3(b+C_{1})(3+5\omega)+\frac{4C_{1}}{\omega}\right]=\omega_{4},\nonumber\\
-\frac{C_{1}}{6}\left[4(C_{1}+b)+3(C_{1}+2b)\omega+\frac{C_{1}}{\omega}\right]=\omega_{5}.\nonumber
\end{eqnarray}

Here the constants of integrations are denoted by $C_i$ for $i =$ 1 to 4. Since the mass function will be zero at centre which indicates that $m_{i}(0)=0$ for $i=$ 1, 2 and hence $C_{2}=C_{4}=0$. We shall get the required approximate analytic mass solution for $\epsilon=1$. So
\begin{equation}
m \approx m_{0}+ m_{1},\label{eq20}
\end{equation}
and finally
\begin{equation}
 m(r) = a_{1}r^{3}+ a_{2}r^{5}+a_{3}r^{7}+a_{4}r^{9}, \label{eq21}
\end{equation}
with $ a_{1}= (C_{1}+C_{3})/3,~a_{2}= (\omega_{5}-b_{1})/5,~a_{3}=\omega_{4}/7$ and $a_{4}=\omega_{3}/9$. There is uncertainty in the sign of the total mass  $m(r)$ as the coefficients $a_2$, $a_3$ and $a_4$ are not zero so their signs are not fixed, so that for the first sight the total mass could appear negative. This issue will be considered in details in the next section.

\section{Strange Star Model with Linear EOS}

\subsection{Density function and constants of integrations}
We write the density function of the star using Eq. (\ref{eq4}) and (\ref{eq21}) as
\begin{equation}
\rho(r)= \frac{1}{4\pi} ( 3a_{1} + 5 a_{2} r^{2} + 7 a_{3} r^{4}+9 a_{4} r^{6} ). \label{eq22} 
\end{equation}

The central density  of star will be 
\begin{equation}
\rho_c=\frac{1}{4\pi}(C_{1}+ C_{3}).\label{eq23}
\end{equation}

In general there is arbitrariness in the values of constants of integrations. So, not all the  values of these constants are important for the  determination of  the central density of a star. Hence, we try to put constrain on the values of constants for a physical star by introducing a model parameter $n$ such that 
\[
C_{1}=\frac{C_{3}}{(n-1)}=\frac{4\pi\rho_{c}}{n}
\]
with $n\geq1$. The above constrain satisfies Eq. (\ref{eq23}). Now it will be possible to find the arbitrary constants of integrations in terms of the physical parameter, such as the central density and the model parameter. Therefore, considering $4\pi\rho_{c}=C$ we get the coefficients	 of the mass function as
\[
 a_{1}= \frac{C}{3},~a_{2}=-\frac{C}{30n}\left[\frac{C}{n}\left(4+3\omega+\frac{1}{\omega}\right)+2b(2+3\omega)\right]-\frac{b^{2}\omega}{10},
 \]
 \[
 a_{3}=\frac{b^{2}\omega}{420}\left[\frac{C}{n}\left(9+15\omega+\frac{4}{\omega}\right)+3b(3+5\omega)\right],~ a_{4}= -\frac{b^{4}\omega^{2}}{540}\left(1+\frac{5\omega}{2}+\frac{1}{2\omega}\right).
 \]

\begin{figure}
\includegraphics[width= 6cm]{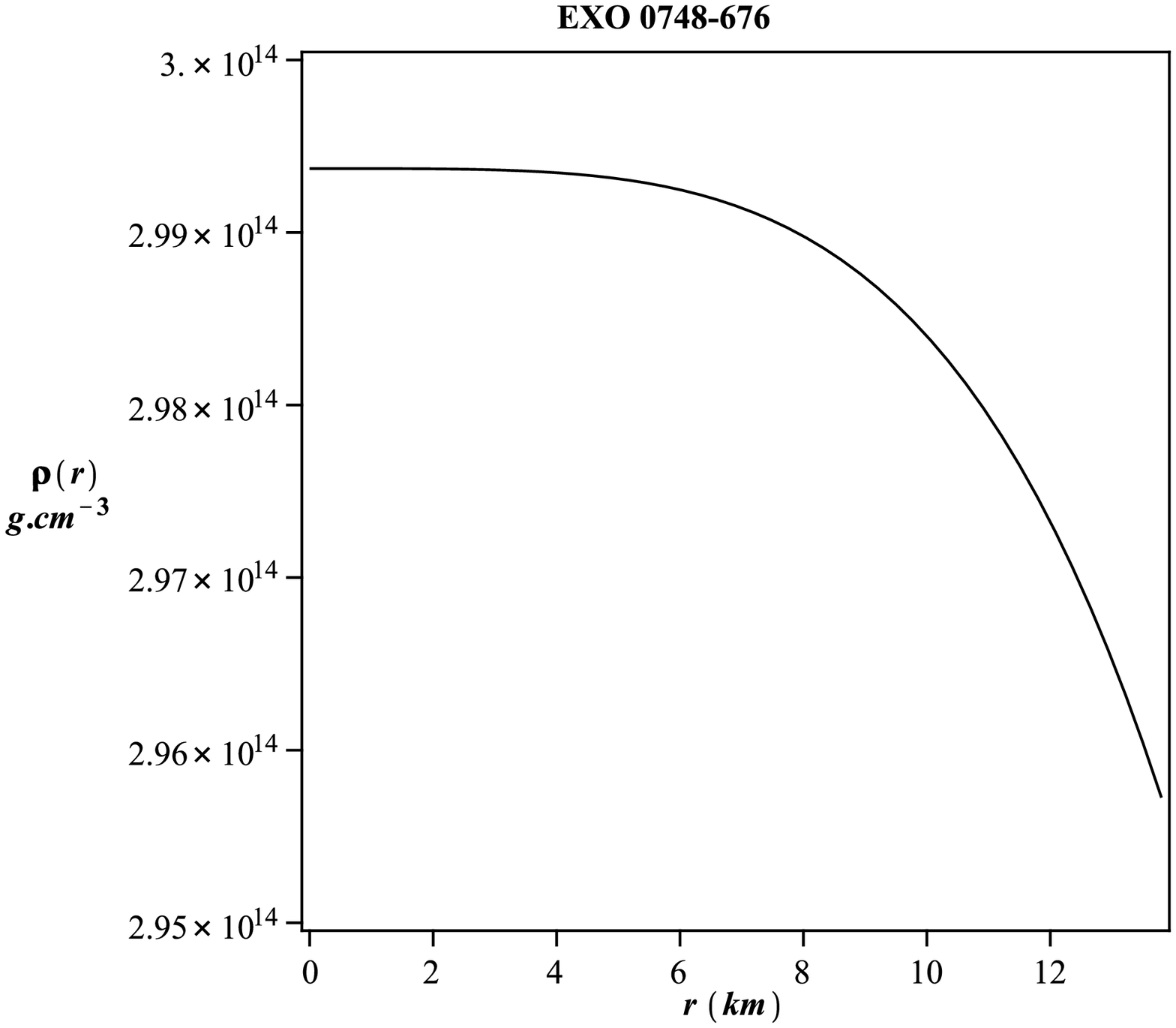}
\includegraphics[width= 6cm]{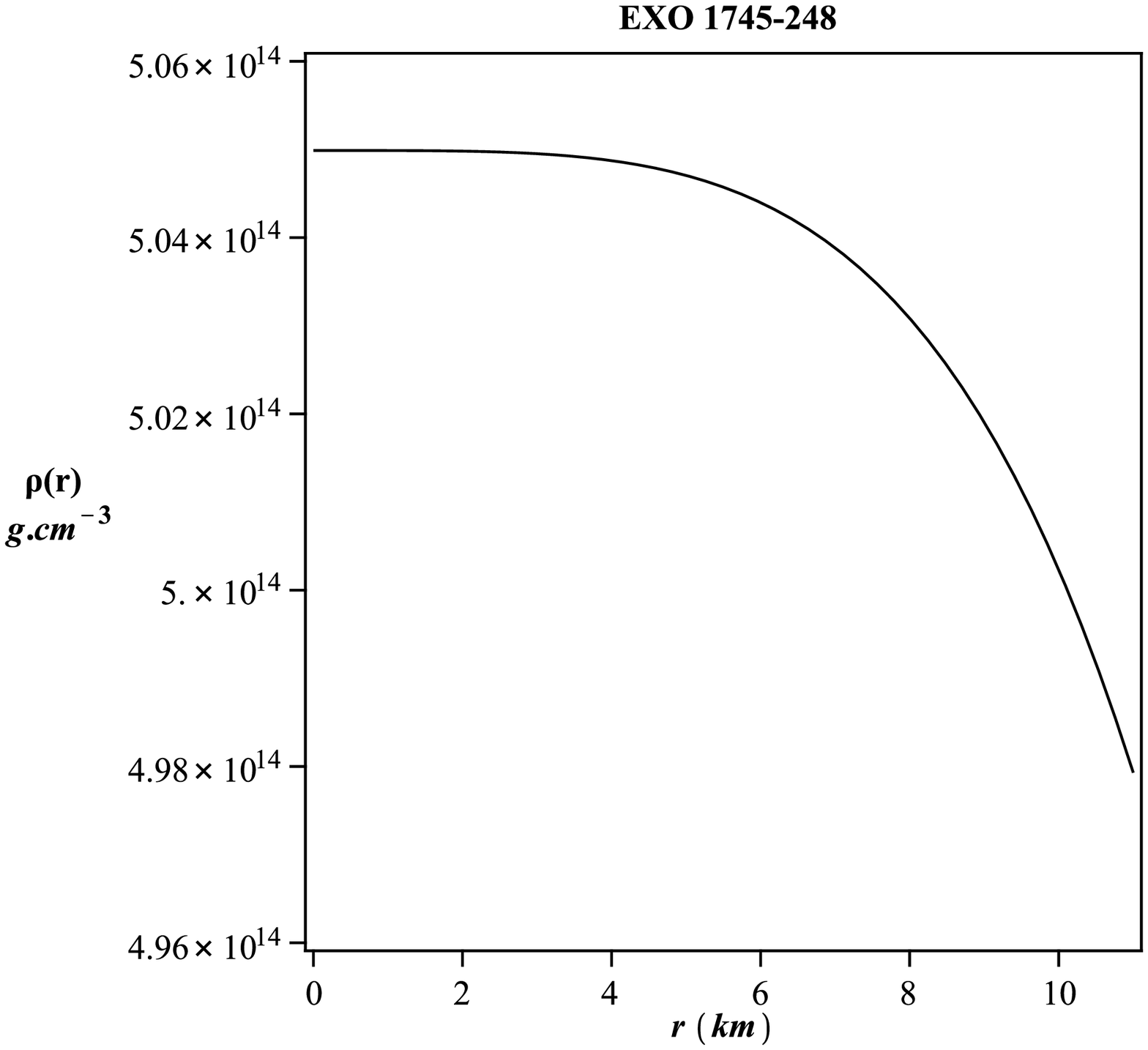}
\includegraphics[width= 6cm]{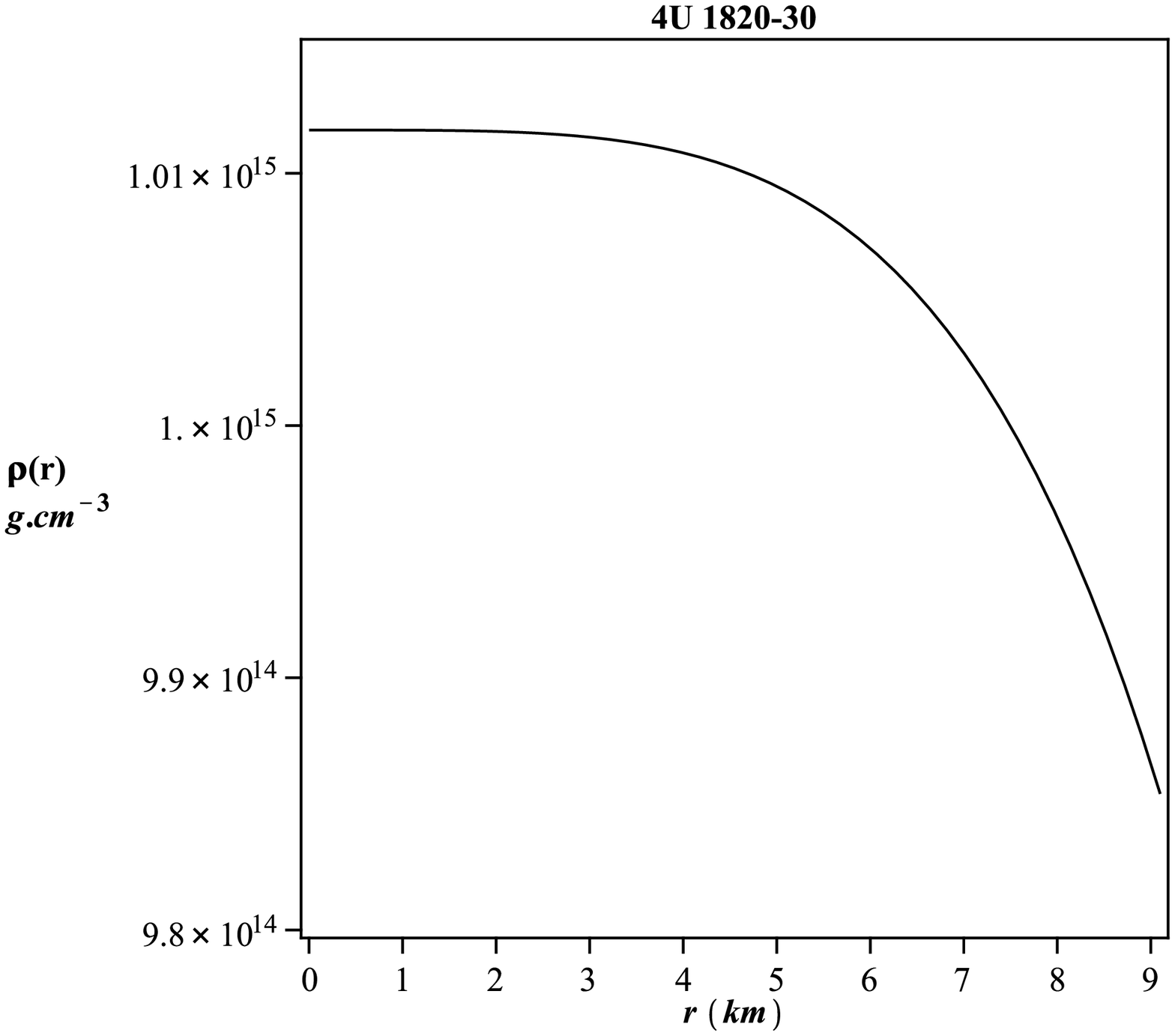}
\includegraphics[width= 6cm]{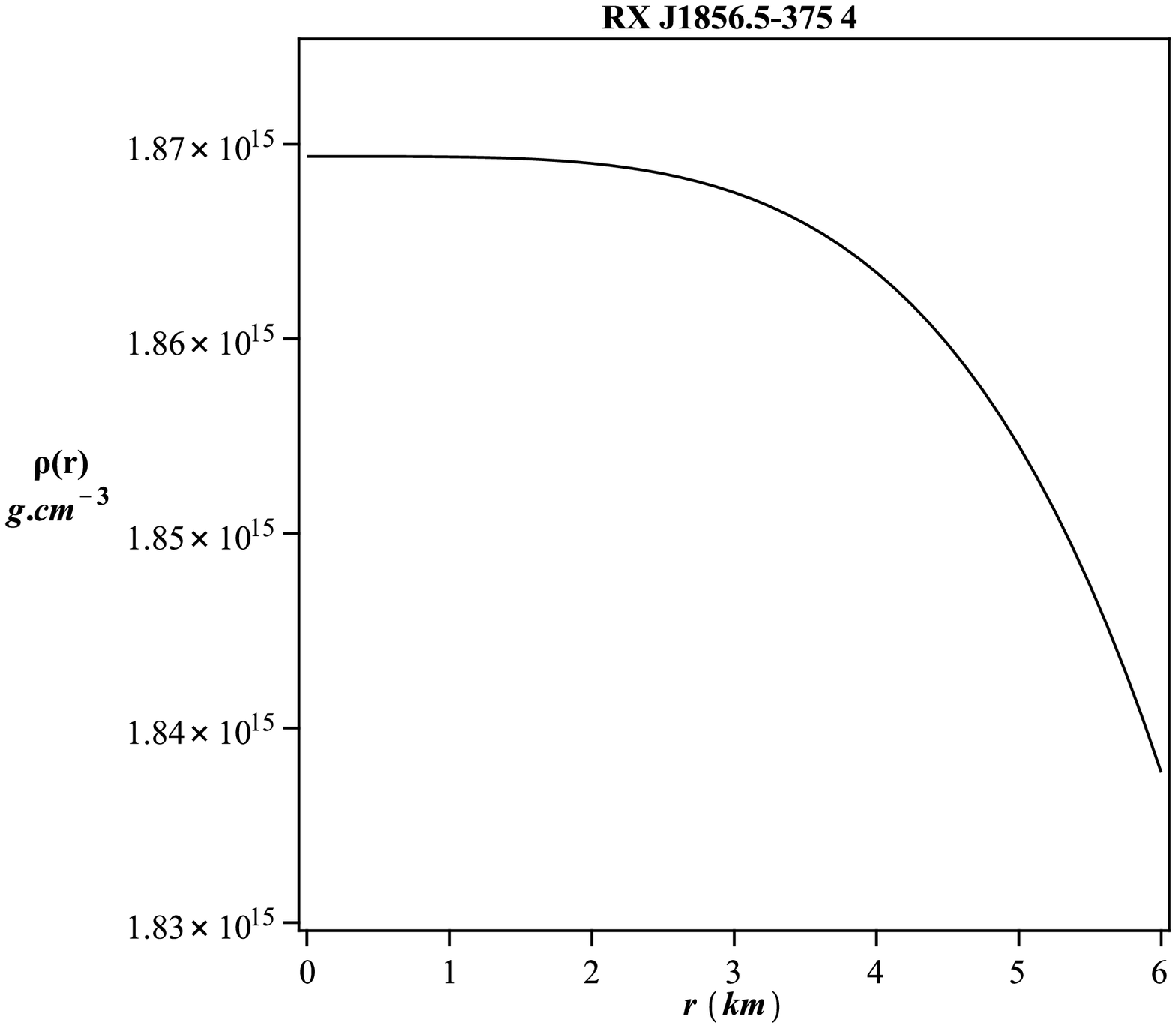}
\caption{Density as a function of radial distance $r$.}\label{Fig1}
\end{figure}

\subsection{Pressure, Radius and Total Mass of the Star} 
We derive the expression for the star using EOS Eq. (\ref{eq3}) and (\ref{eq22}) as
\begin{equation}
p(r)= \frac{\omega}{4\pi} ( 3a_{1} + 5 a_{2} r^{2} + 7 a_{3} r^{4}+9 a_{4} r^{6} )+\frac{\omega b}{4\pi}. \label{eq24}
\end{equation}

\begin{figure}
\centering
\includegraphics[width= 6cm]{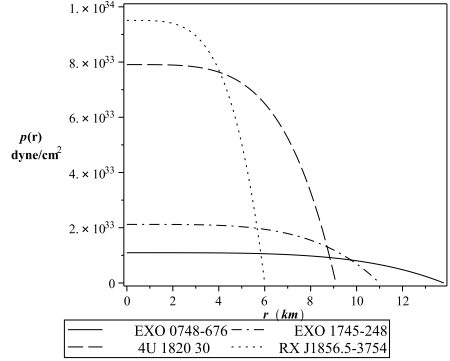}
\caption{Pressure as a function of radial distance $r$.}\label{Fig2}
\end{figure}

The pressure of a star decreases from the core to the surface as such it is a decreasing function with respect to radial distance. At each radial distance the pressure acts against collapse of the stars due to gravity. Since the stabilization process occurs from centre to the surface of star,  we can assume the condition $P(R)=0$ at boundary of star. Under this boundary condition of vanishing pressure $b$ yields as $-4\pi\rho(R)$. Hence, we get the following expression 
\begin{equation}
C+b+5a_{2}R^2+7a_3R^4+9a_4R^6= 0.\label{eq25}
\end{equation}

By solving Eq. (\ref{eq25}), we can get radius of the star which obviously depends on different values of $(\omega,n)$. Hence, the total mass of the star 
\begin{equation}
M = 4\pi\int_{0}^{R}\rho(r)r^{2}dr, \label{eq26}
\end{equation}
reduces to
\begin{equation}
M = a_{1}R^{3}+ a_{2}R^{5}+a_{3}R^{7}+a_{4}R^{9}.\label{eq27}
\end{equation} 

As it can be observed that the positivity of the total mass depends on values of the coefficients which will be determined later on in Section 5.

\subsection{Causality and Stability Condition}
The expression for the acoustic speed in the perfect fluid system is given by
\begin{equation}
v= \sqrt{\frac{dp}{d\rho}}= \sqrt{\omega}.\label{eq28}
\end{equation}

A solution will be causal if the acoustic speed must be less than the light speed, i.e., $0 < v^{2} < 1$ \cite{Herrera1992}.
The definition of adiabatic index for isotropic stellar structure~\cite{Chandrasekhar1964,Merafina1989} is 
\begin{equation}
\Gamma= \left(\frac{\rho+p}{p}\right)\frac{dp}{d\rho}=\omega+\left(1-\frac{16\pi B}{( 3a_{1} + 5 a_{2} r^{2} + 7 a_{3} r^{4}+9 a_{4} r^{6} )}\right)^{-1}. \label{eq29}
\end{equation}

For the stable stellar structure we will always have $\Gamma> \frac{4}{3}$ \cite{Chandrashekhar1964a,Bondi1964,Wald1984}.

\subsection{Time-time component of the metric}
The unknown time-time component of metric tensor in the line element can be derivable from Eq. (\ref{eq5}) using Eq. (\ref{eq3}) and (\ref{eq4}). It takes the following form   
\begin{equation}
g_{tt}(r) =  g_{tt}(0) \frac{e^{I(r)}}{\left(1-\frac{2m}{r}\right)^{\omega}}, \label{eq30}
\end{equation}
where 
\[
I(r)=\int \frac{\left[(1+\omega)\frac{2m}{r^2}+\omega b r\right]}{\left(1-\frac{2m}{r}\right)}dr.
\]

We can obtain the interior solution of Einstein's field equations using the line element (2). The exterior solution, which is known as the Schwarzschild  metric, was obtained by Schwarzschild himself in 1916 as follows
\begin{equation}
ds^{2}=-\left(1-\frac{2M}{R}\right) dt^{2}+\left(1-\frac{2M}{R}\right)^{-1}dr^{2}+r^{2} d\theta^{2}+ r^{2} {sin}^{2}\theta d\phi^{2}.
\end{equation}

So, for physical validity the interior metric (2) will be equal to  the Schwarzschild metric at the boundary of the star. Thus we get
\begin{equation}
g_{tt}(R)= \left(1- \frac{2M}{R}\right). \label{eq31}
\end{equation}

Using the above mentioned matching condition we get 
\begin{equation}
g_{tt}(0)= \left(1-\frac{2M}{R}\right)^{\omega+1}e^{-I(R)}.\label{eq32}
\end{equation}

As $g_{tt}(0)$ is finite and hence we have non-singular time-time metric component component of metric tensor.

\begin{figure}
\includegraphics[width= 6cm]{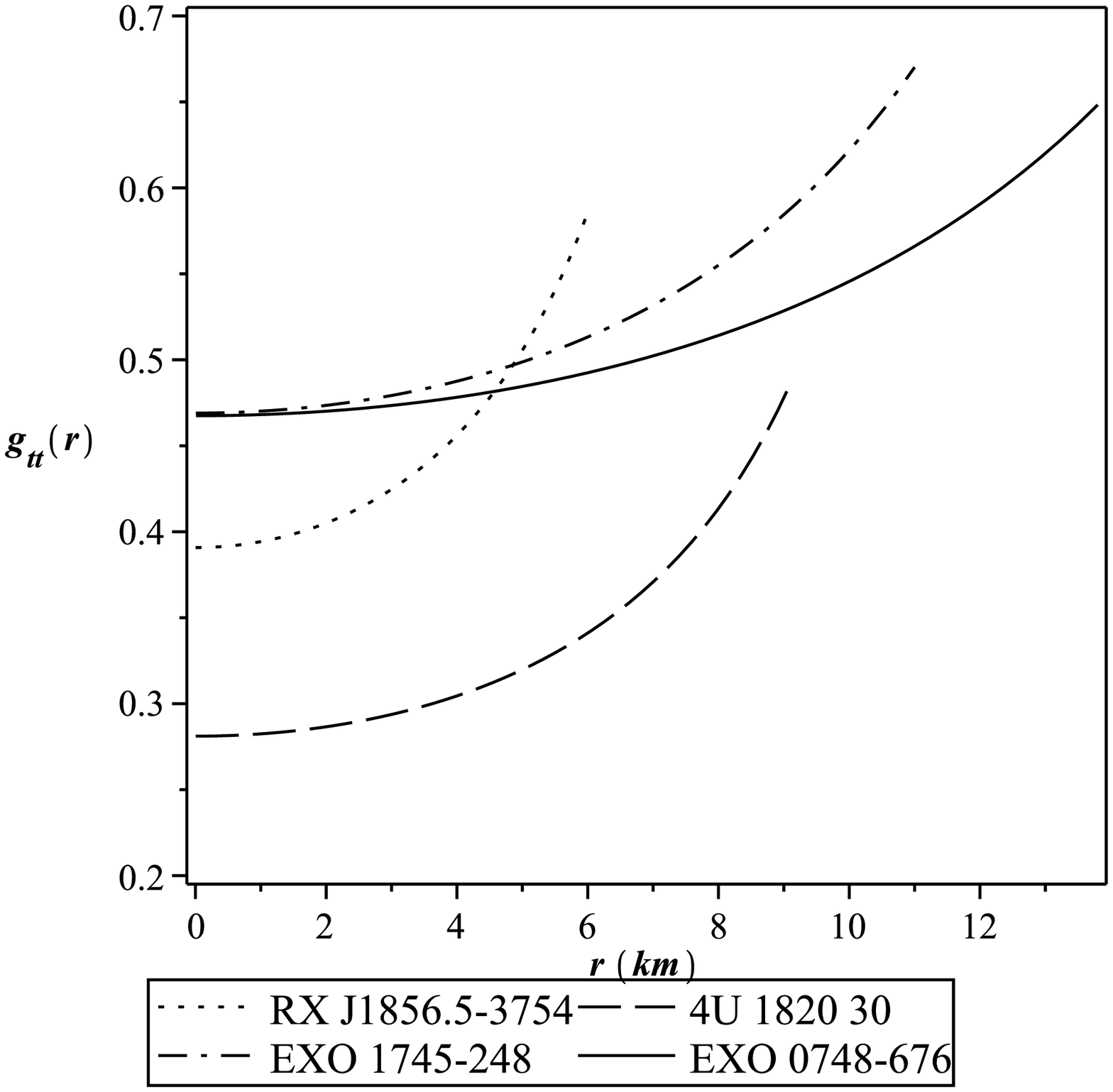}
\includegraphics[width= 6cm]{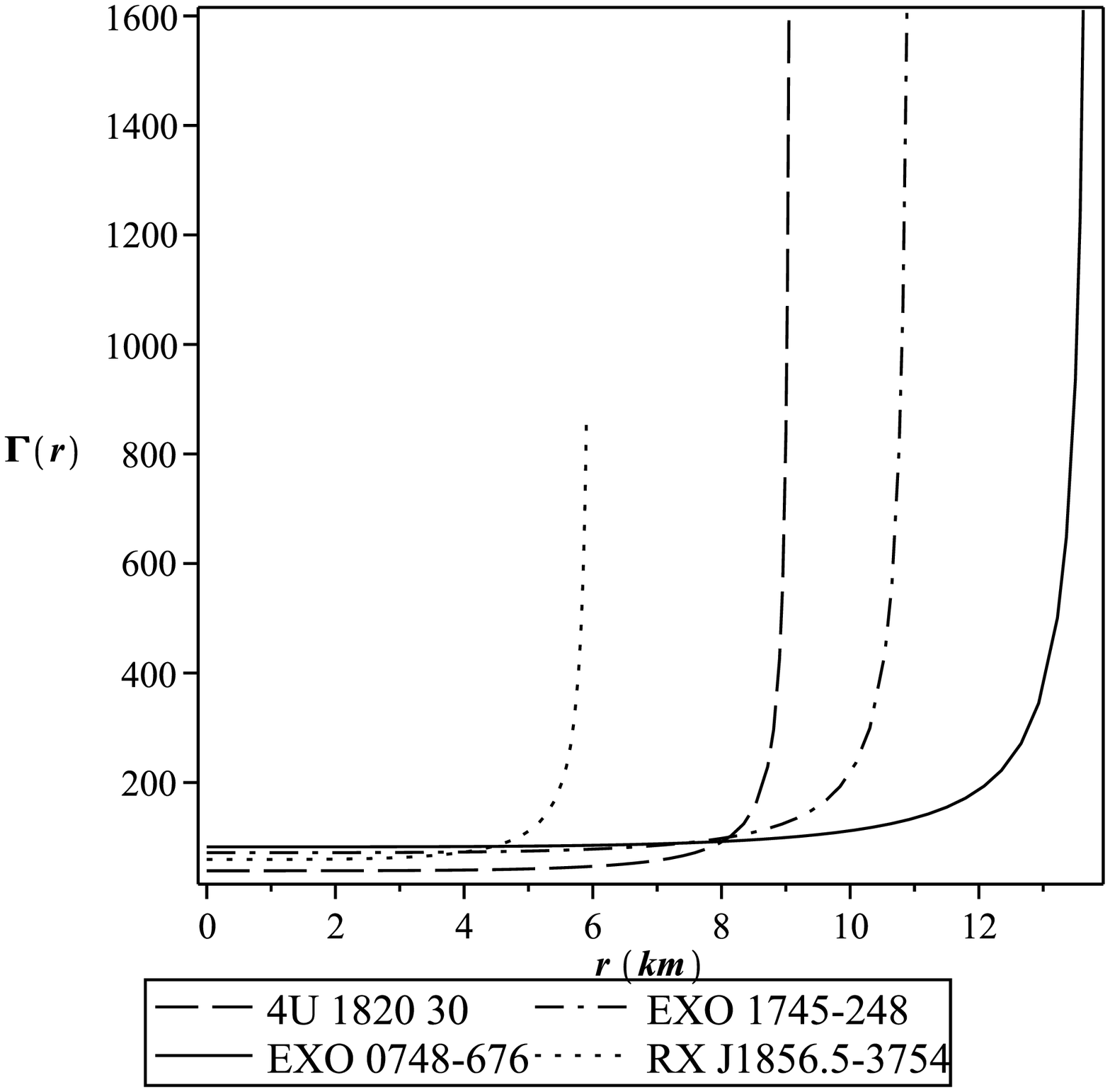}
\caption{ time-time metric component and adiabatic index as a function of radial distance $r$s.}\label{Fig3}
\end{figure}

\subsection{Compactness and Redshift}
The measurement of compactness of a star is given by ratio of Schwarzschild radius $(R_{S}= 2GM/c^{2})$ and radius of star $(R)$. Here compactification factor of the star is defined  as $u=m(r)/r$ which can be written as
\begin{equation}
u(r)=a_{1}r^{2}+ a_{2}r^{4}+a_{3}r^{6}+a_{4}r^{8}. \label{eq33}
\end{equation}
The total mass to radius ratio, i.e., $u(R)=M/R$  is given by
\begin{equation}
\frac{M}{R}=a_{1}R^{2}+ a_{2}R^{4}+a_{3}R^{6}+a_{4}R^{8}. \label{eq34}
\end{equation}

A stable and physical stellar structure  $u(R)$ should be always less than 4/9. This is known as Buchdahl's condition~\cite{Buchdahl1959}.  
 
We can define the redshift function~\cite{Ozel2013} in the internal region of a star as 
\begin{equation}
Z(r)+1= \frac{1}{\sqrt{g_{tt}(r)}}= [1-2u(R)]^{-\frac{1}{2}}.\label{eq36}
\end{equation}

Using Buchdahl's condition, i.e., $u(R)\leq 4/9$ it can be shown that $Z_s\leq2$. Substituting Eq. (34) in Eq. (36) we therefore get the expression for the surface redshift as
\begin{equation}
Z_{s}+1= [1-2(a_{1}R^{2}+ a_{2}R^{4}+a_{3}R^{6}+a_{4}R^{8})]^{-\frac{1}{2}}.\label{eq37}
\end{equation}

\begin{figure}
\includegraphics[width= 6cm]{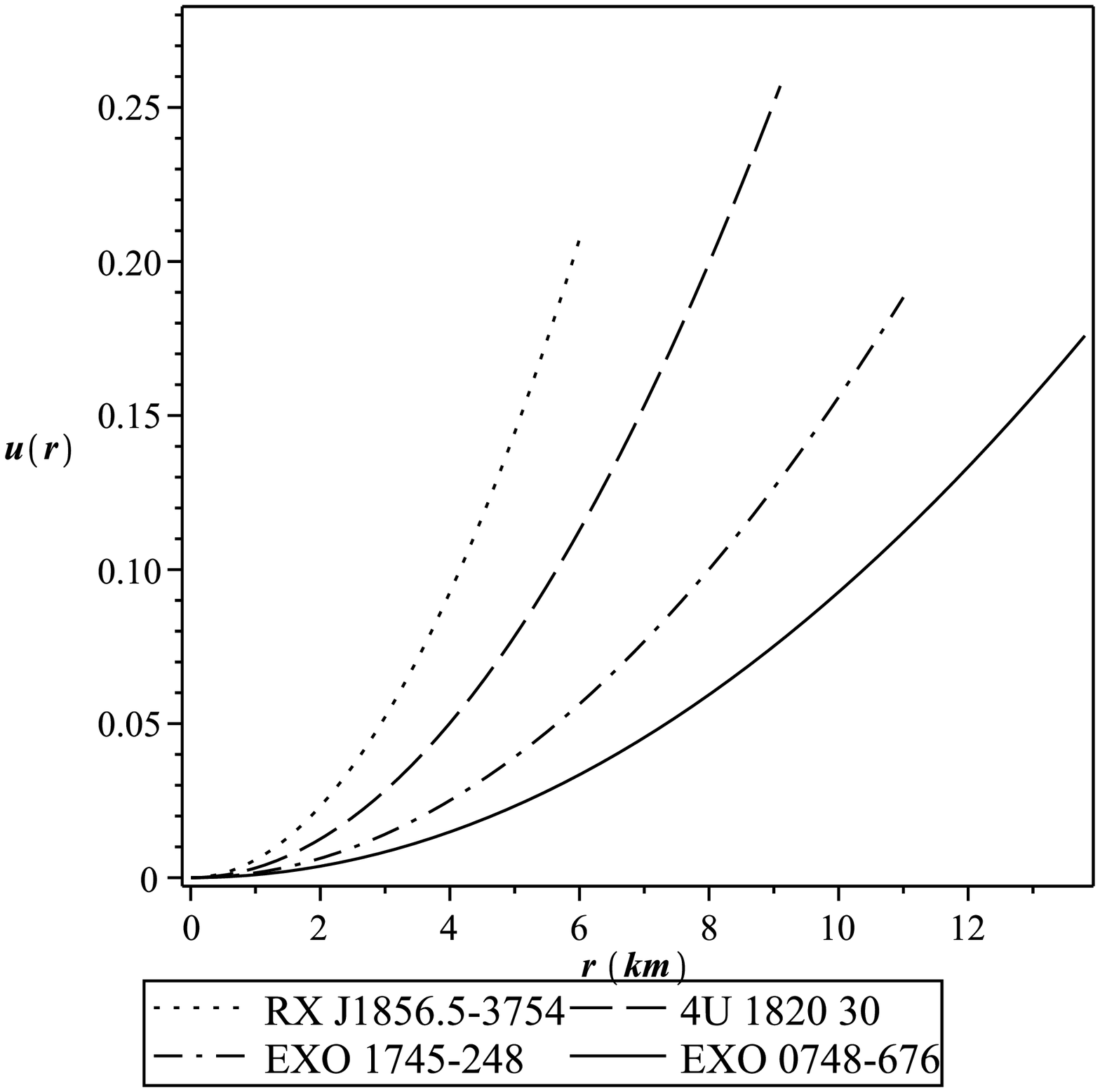}
\includegraphics[width= 6cm]{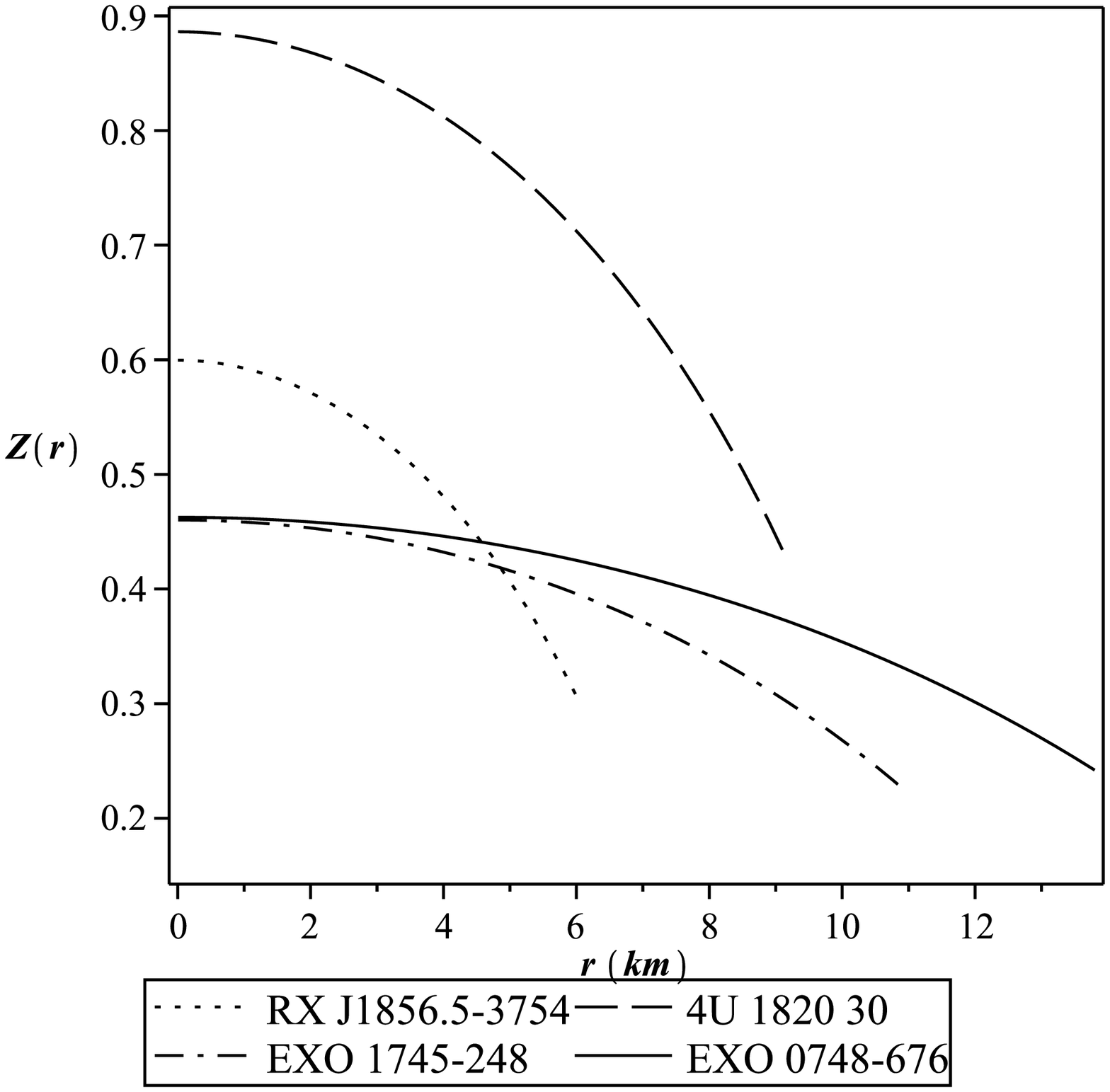}
\caption{Compactness and redshift as a function of radial distance $r$.}\label{Fig4}
\end{figure}

\subsection{Energy Conditions} 
The components of energy-momentum tensor are expected to satisfy energy conditions. Let us consider $k^n$ to be null vector whereas $\xi^a$ and $\eta^a$ to be timelike vector. There are different energy conditions which can be expressed in the following manner:  
\[
(i)~Null~energy~condition~(NEC):~T_{mn}k^mk^n\geq0,
\]

\[
(ii)~Weak~ energy ~condition~(WEC):~T_{mn}\xi^m\xi^n\geq0,
\]

\[
(iii)~Strong ~energy~ condition~(SEC):~\left(T_{mn}-\frac{1}{2}Tg_{mn}\right)\xi^m\xi^n\geq0,
\]

\[
(iv)~Dominant~ energy ~condition~(DEC):
1.~T_{mn}\xi^m\xi^n\geq0~and~T^a_{~n}\xi^n~is~causal,
\]
\[
2.~For~two~cooriented~vectors~\xi~and~\eta,~T_{mn}\xi^m\eta^n\geq0.
\]

The implication of WEC is that the total energy density will be positive for an observer travelling along the time-like geodesic and the pressures will be positive in the space-like directions, SEC suggests that a local observer following the time-like curve will see the gravity as an attractive force and NEC is satisfied whenever WEC is satisfied. However SEC does not imply WEC whereas DEC implies WEC.  In our model for isotropic fluid these conditions reduced to effective form as
\begin{eqnarray}
(i)~NEC: \rho+p\geq 0,\nonumber\\
(ii)~WEC: \rho+p\geq 0,~\rho\geq 0,\nonumber\\
(iii)~SEC: \rho+p\geq 0,~\rho+3p\geq 0,\nonumber\\
(iv)~DEC: \rho\geq 0, ~ \rho> \mid p \mid. \nonumber
\end{eqnarray}

The validation and violation of energy conditions lead to many physical phenomena, such as formation of the event horizon~\cite{Penrose1965}, formation of singularities~\cite{Penrose1965}, black hole thermodynamics~\cite{Hawking1971} etc. So, we shall check whether the energy conditions are satisfied in the present model for its physical validation.

\begin{figure}
\includegraphics[width= 6cm]{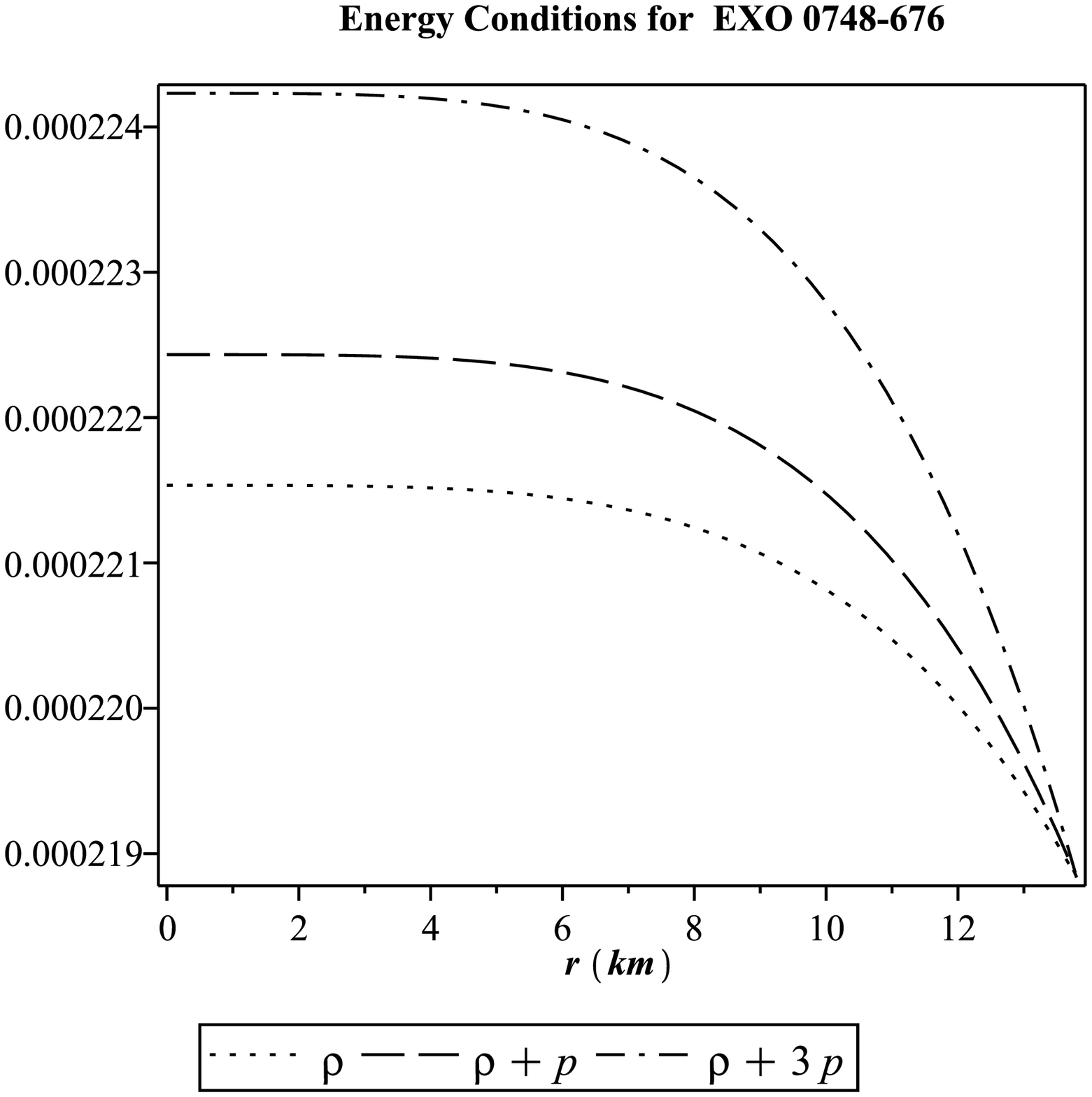}
\includegraphics[width= 6cm]{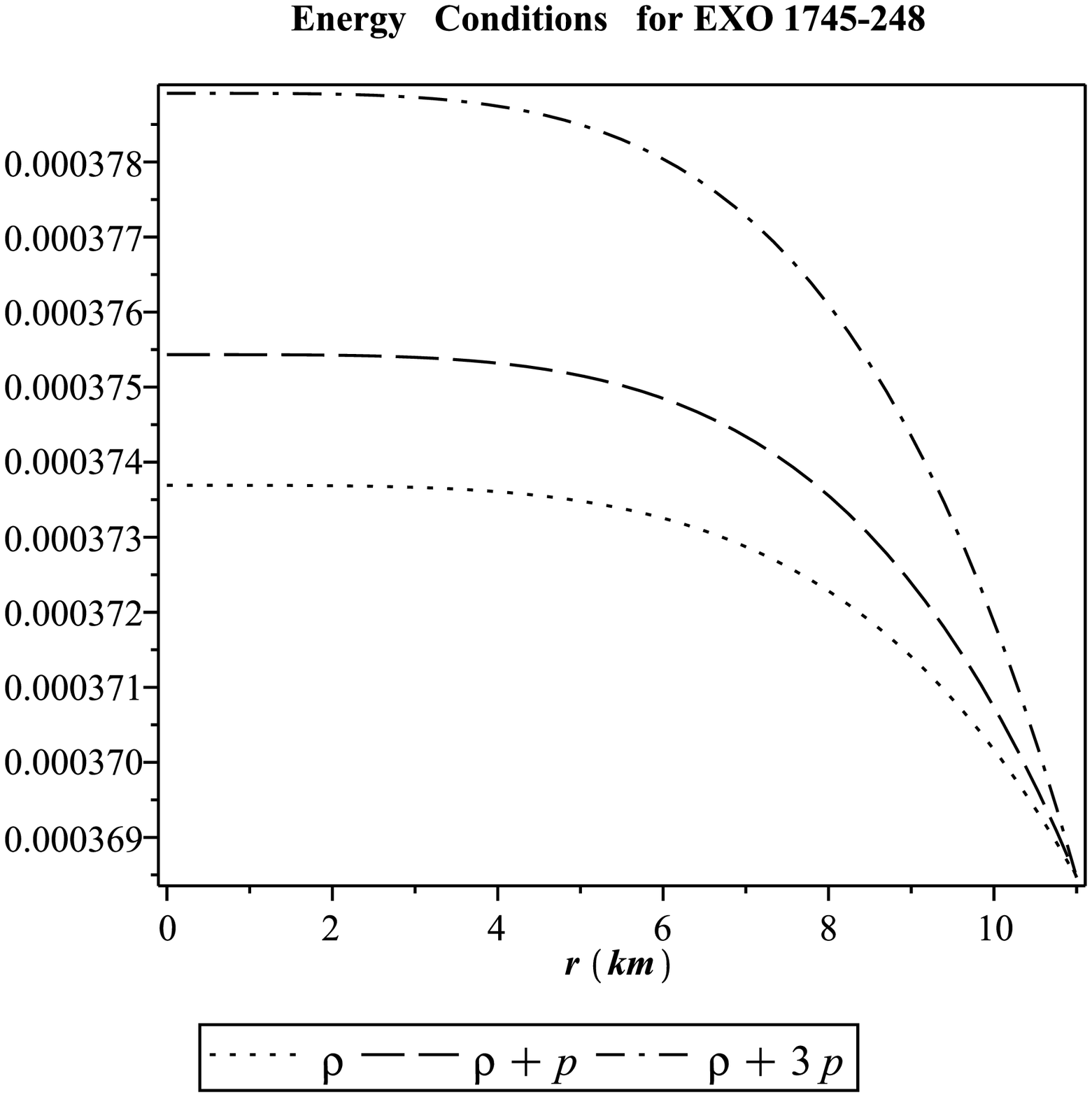}
\includegraphics[width= 6cm]{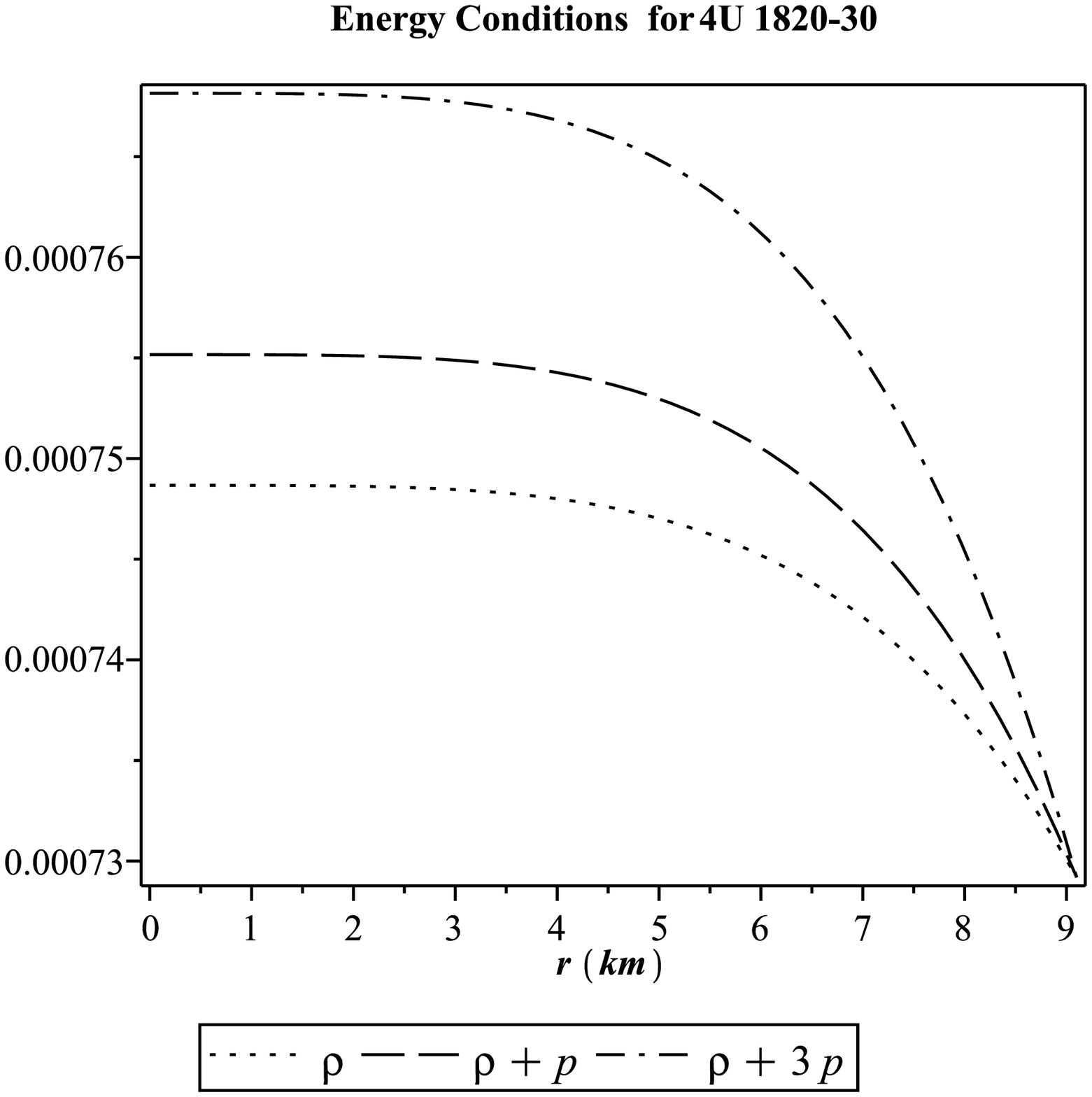}
\includegraphics[width= 6cm]{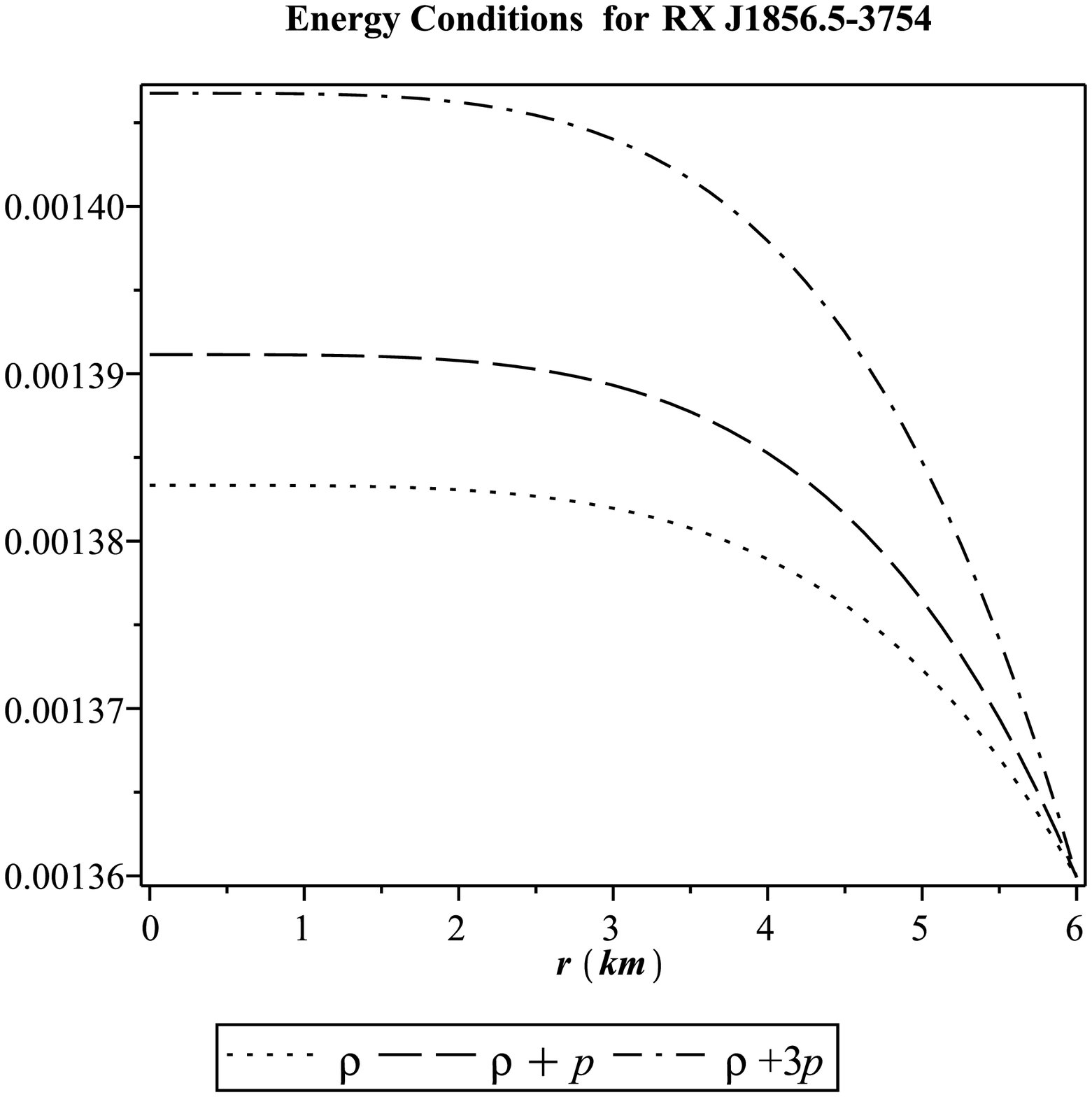}
\caption{Energy condition as a function of radial distance $r$.}\label{Fig5}
\end{figure}

\section{MIT Bag Model for Strange Star Candidates}
The EOS for quark matter in MIT bag model~\cite{Chodos1974} is given by
\begin{equation}
p = \frac{1}{3} (\rho-4B). \label{eq38}
\end{equation}

So, comparing with EOS (\ref{eq3}) we will have $\omega=\frac{1}{3}$, $b=-16\pi B$ respectively. The coefficients in the mass function are as follows
\[
 a_{1}= \frac{C}{3},~ a_{2}=-\frac{4}{15}\left(\frac{C}{n}-4\pi B\right)\left(\frac{C}{n}-8\pi B\right),
 \]
 \[
 a_{3}=\frac{8}{315}(4\pi B)^{2}\left(\frac{13C}{n}-112\pi B\right),~ a_{4}= -{\frac {128}{729}}(4\pi{B})^{4}.
 \]   

Substituting the above coefficients in Eq. (\ref{eq25}) we can get an algebraic equation for radius of a star. So the solution for radius will depend on the physical parameters $B$, $\rho_c$ and model parameter $n$. For physical solutions the parameters shall have constrained range of values. We define a dimensionless parameter as $x =4\pi B R^{2}$ which reduces Eq. (\ref{eq25}) into a cubic equation for $x$. We at first try to evaluate general solutions of the cubic equation in terms of discriminant and coefficients. Due to arbitrariness of the parameters ($B$,~$\rho_c$,~$n$) we see that in most of the cases the roots of the cubic equations are either imaginary or violate the Buchdahl condition. It is to note that using Cardano's formula for solution of cubic equation, we failed to find realistic solutions for physical radii of stars. Therefore to find physical solution we need to know the allowed ranges of the parameters satisfying realistic conditions.  

We see that $\rho(R)=4B$ from Eq. (\ref{eq37}). We parametrize the bag constant using model parameter $n$ and central density as free parameter. Since for any physical stellar system the central density is always greater than the surface density we take a linear relation as  $\rho_c=n B$ where $n>4$. Under this consideration the compactness of the star is given by 
\begin{equation}
\frac{M}{R}=U(x)= \frac{n}{3}x-\frac{8}{21}x^{3}-\frac{128}{729}x^{4}. \label{eq39}
\end{equation}

From Eq. (\ref{eq25}) we will get the relation between $n$ and $x$ as 
\begin{equation}
n=4+\frac{8}{3}x^{2}+\frac{128}{81}x^{3}. \label{eq40}
\end{equation}

Substituting Eq. (\ref{eq40}) in Eq. (\ref{eq39}) we get compactness as function of $x$, 
\begin{equation}
U(x)= \frac{4}{3}x\left(1+\frac{8}{21}x^{2}+\frac{64}{243}x^{3}\right).\label{eq41}
\end{equation}

A stable and physical star must follow Buchdahl's condition, we should always have $x<0.3183388920$ and $n<4.32$ . This constraint gives the upper bound for bag constant for the star of radius $R$ as $B< (0.025/R^2)$.  

\begin{figure}
\centering
\includegraphics[width=8cm]{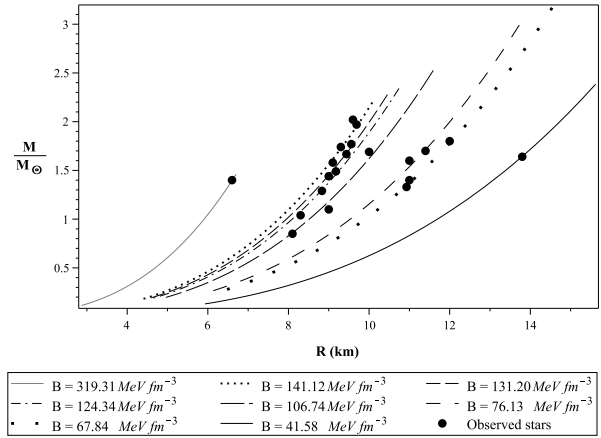}
\caption{Mass-radius relation.}\label{Fig6}
\end{figure}

\section{Physical Validity of the Model}
To test the physical validity of the star model and to determine possible definite range of values of bag constants we study 20 compact stars considering as strange stars. Now we discuss about the stars very briefly. The masses of $Vela~X-1$, $LMC~X-4$, $CEN~X-3$ are $1.77 \pm 0.08~M_\odot$, $1.29 \pm 0.05~M_\odot$ and $1.49 \pm 0.08~M_\odot$~(Rawls et al. 2011). $PSR~J1903+0327$ of mass $1.667\pm0.021~M_\odot$ ruled out sofest EOS~(Freire et al. 2011). Gangopadhyay et al.~(2013) predicted radii of these six pulsars considering EOS compatible to strange stars. Li et al.~(1995) first proposed that $Her~X-1$ could be strange star. However the star is NS as supported by other works~(Reynolds et al. 1997, Madsen 1997). Dey et al.~(1998) gives a strange star model using phenomenological density dependent scalar potential for $Her~X-1$. Using Shaprio delay the mass of binary milisecond $PSR~J1614-2230$ is precisely determined to be $1.97 \pm 0.04~M_\odot$~(Demorest et al. 2011) which support presence quark matter in the NS (Alford et al. 2007, Kurkela et al. 2010, {\"O}zel et al. 2010). Other strange star candidates are $PSR~1937+21$~(Ren-Xin et al. 2001), $EXO~0748-676$~(Xu 2003, Alford et al. 2007), $4U~1636-536$~(Sinha et al. 2002), $KS~1731-260$~(Sinha et al. 2002), $4U~ 1820-30$~(Dey et al. 1998, Sinha et al. 2002, Bombaci 1997), $4U~1728-34$~(Li et al. 1999), $SAX~J1808.4-3658$~(Li et al. 1999a, Bombaci 2000) and $RX~ J185635-3754$~(Kohri et al.2003a, Kohri et al.2003b, Drake et al. 2002, Xu 2002, Prakash et al. 2003, Henderson and Page 2007). We have taken the mass and radius of NS, e.g. $4U~1822-371$~(Iaria et al. 2015), $4U~1608-52$~(G{\"u}ver et al. 2010), $EXO~1745-248$~(G{\"u}ver et al. 2009), $Cygnus~X-2$~(Titarchuk and Shaposhnikov 2002), $SAX~J1748.9-2021$~(G{\"u}ver and {\"O}zel 2013). We prepare a data sheet in Table 1 for all the above discussed compact stars.

The mass-radius relation is shown in Fig. \ref{Fig6}. We plot maximum mass and corresponding bag constants using the data from Table 1 in Fig. \ref{Fig6}. From this figure we note that as the value of $B$ increases correspondingly stiffness of the M-R curves increases. One can find that the lower bound of bag constant is 41.58~MeV~fm$^{-3}$ whereas the upper bound is 319.31~MeV~fm$^{-3}$. The lower and upper bounds of $B$ are seen to be less probabilistic with respect to observed stars. As a consequence extremely stiff EOS and extremely soft EOS are less probable. However we note from the mass-radius curves in Fig. \ref{Fig6} that most probable values of bag constant with respect to observed stars lie in the range 67.84~MeV~fm$^{-3}< B <$141.12~MeV~fm$^{-3}$. So, intermediate EOS are seen to be physically more acceptable to describe different features of strange stars.

To show the physical validation of properties of stars we take four stars such as $EXO~0748-676$, $EXO~1745-248$, $RX~J185635-3754$ and $4U~1820-30$. The variation of the density, pressure, compactness, redshift, time-time component of metric, adiabatic index and energy conditions are shown in Figs. \ref{Fig1}-\ref{Fig5}. 

\begin{sidewaystable*}[htbp!]
\vspace{12.0cm}
\centering
\resizebox{\columnwidth}{!}{
\begin{tabular}{@{}cccccccccc@{}}
\hline\hline
 Star & M & R & $B$ & M/R & $Z_s$ & $Z_c$ & $\rho_c \times 10^{15}$ & $p_c \times 10^{33}$  \\ 
   
   &  ($M_{\odot}$) &  (km) & (MeV fm$^{-3})$ & &  &  & (gmcm$^{-3})$ & (dynecm$^{-2})$ \\ \hline \hline \\
 
 4U 1636-536 \cite{Kaaret1997} & $2.02\pm0.12$ & $9.6\pm0.6$ & $149.77^{-17.47}_{+21.17}$ & $0.31\mp0.00$ & $0.63^{-0.01}_{+0.00}$ & $1.36\mp0.01$ & $1.11^{-0.13}_{+0.15}$ & $12.63^{-1.54}_{+1.88}$  \\ \hline
 
 PSR J1614-2230 & $1.97\pm0.04$\cite{Demorest2011} & $9.69\pm0.2$\cite{Gangopadhyay2013} & $142.24^{-5.73}_{+6.11}$ & $0.30\mp0.00$ & $0.58\mp0.00$ & $1.25\mp0.00$ & $1.05\mp0.04$ & $11.18^{-0.45}_{+0.49}$  \\ \hline
 
 KS 1731-260 \cite{Abubekerov2008} & 1.8 & 12 & 69.11 & 0.22 & 0.34 & 0.68 & 0.50 & 2.93  \\ \hline
 
 Vela X-1 & $1.77\pm0.08$\cite{Rawls2011} & $9.56\pm0.08$\cite{Gangopadhyay2013} &  $133.58^{+2.41}_{-2.61}$ & $0.27\pm0.01$ & $0.49\pm0.03$ & $0.98\pm0.02$ &  $8.69^{+0.82}_{-0.80}$ & $1.01\pm0.08$  \\ \hline
 
 4U 1608-52 \cite{Poutanen2003}& $1.74\pm0.14$ & $9.3\pm1.0$ & $142.58^{-29.08}_{+41.65}$ & $0.28\mp0.01$ & $0.50^{-0.02}_{+0.03}$ & $1.04^{-0.05}_{+0.07}$ & $1.04^{-0.21}_{+0.31}$ & $9.47^{-1.76}_{+3.52}$  \\ \hline
 
 RX J1856.5-3754\cite{Prakash2003} & $1.7\pm0.4$ & $11.4\pm2$ & $76.13^{-18.29}_{+27.89}$ & $0.22^{+0.01}_{-0.02}$ & $0.34^{+0.03}_{-0.04}$ & $0.66^{+0.06}_{-0.08}$ & $0.55^{-0.13}_{+0.20}$ & $3.18^{-0.50}_{+0.55}$  \\ \hline
 
4U 1822-37  \cite{Iaria2015}& $1.69\pm0.13$ & 10 & $111.76^{+8.32}_{-8.36}$ & $0.25\pm0.02$ & $0.41^{+0.06}_{-0.05}$ & $0.84^{+0.14}_{-0.12}$ & $0.82^{+0.06}_{-0.07}$ & $6.01^{+1.53}_{-1.26}$  \\ \hline
 
 PSR J1903+0327 & $1.667\pm0.021$ \cite{Freire2011} & $9.438\pm0.03$ \cite{Gangopadhyay2013}  & $131.20^{+0.36}_{-0.37}$ & $0.26\pm0.00$ & $0.45\pm0.01$ & $0.92^{+0.03}_{-0.02}$ & $0.96\pm0.00$ & $7.76\pm0.17$  \\ \hline
 
 EXO 0748-676 \cite{Medin2014} & $1.64\pm0.38$ & $13.8^{+0.6}_{-2.0}$ & $41.58^{+3.36}_{+9.59}$ & $0.18^{+0.03}_{-0.02}$ & $0.24^{+0.07}_{+0.03}$ & $0.46^{+0.14}_{+0.05}$ & $0.30^{0.02}_{+0.07}$ & $1.09^{+0.57}_{-0.01}$ \\ \hline
 
 SAX J1808.4-3658 \cite{Poutanen2003}& 1.6 & 11 & 79.80 & 0.22 & 0.32 & 0.64 & 0.58 & 3.18  \\ \hline
 
 4U 1820-30 \cite{Guver2010} & $1.58\pm0.06$ & $9.1\pm0.4$ & $138.55^{-12.13}_{+13.95}$ & $0.26\mp0.00$ & $0.43^{-0.00}_{+0.01}$ & $0.89\mp0.01$ & $1.01^{-0.09}_{+0.10}$ & $7.91^{-0.78}_{+0.91}$ \\ \hline
 
 Cen X-3 & $1.49\pm0.08$ \cite{Rawls2011}& $9.178\pm0.13$\cite{Gangopadhyay2013} & $127.60^{+1.16}_{-1.43}$ & $0.24\pm0.01$ & $0.39^{+0.02}_{-0.03}$ & $0.78\pm0.06$ & $0.92\pm0.02$ & $6.35^{+0.57}_{-0.58}$ \\ \hline

 Cyg X-2\cite{Titarchuk2002} & $1.44\pm0.06$ & $9.0\pm0.5$ & $130.83^{-14.91}_{+17.95}$ & $0.24^{-0.01}_{+0.00}$ & $0.38\mp0.02$ & $0.76\mp0.02$ & $0.95^{-0.11}_{+0.13}$ & $6.32^{-0.87}_{+1.08}$ \\ \hline
 
 PSR 1937+21\cite{Ren-Xin2001} & 1.4 & 6.6 & 319.31 & 0.31 & 0.64 & 1.39 & 2.32 & 27.36  \\ \hline
 
 EXO 1745-248 \cite{Ozel2009} & 1.4 & 11 & 70.01 & 0.19 & 0.27 & 0.51 & 0.50 & 2.12  \\ \hline
 
 SAX J1748.9-2021 \cite{Guver2013}& $1.33\pm0.33$ & $10.93\pm2.09$ & $67.84^{-17.78}_{+28.67}$ & $0.18\pm0.02$ & $0.25\pm0.02$ & $0.48^{+0.04}_{-0.05}$ & $0.49^{-0.12}_{+0.20}$ & $1.87^{-0.35}_{+0.43}$ \\ \hline
 
 LMC X-4 & $1.29\pm0.05$\cite{Rawls2011} & $8.831\pm0.09$ \cite{Gangopadhyay2013}& $124.34^{+0.87}_{-1.01}$ & $0.22^{+0.00}_{-0.01}$ & $0.33^{+0.03}_{-0.02}$ & $0.64^{+0.04}_{-0.03}$ & $0.90\pm0.01$ & $5.00^{+0.32}_{-0.34}$  \\ \hline
 
4U 1728-34 \cite{Li1999}& 1.1 & 9 & 100.50 & 0.18 & 0.25 & 0.48 & 0.72 & 2.80  \\ \hline 
 
 SMC X-1 & $1.04\pm0.09$ & $8.301\pm0.2$ & $121.05^{+1.28}_{-1.97}$ & $0.18\pm0.01$ & $0.26\pm0.02$ & $0.50\pm0.05$ & $0.87^{+0.01}_{-0.02}$ & $2.95^{+1.10}_{-0.90}$ \\ \hline
 
Her X-1 & $0.85\pm0.15$\cite{Abubekerov2008} & $8.1\pm 0.41$ \cite{Gangopadhyay2013} & $106.74^{+1.39}_{-3.87}$ & $0.16^{+0.01}_{-0.03}$ & $0.20^{+0.04}_{-0.03}$ & $0.38^{+0.07}_{-0.06}$ & $0.77^{+0.01}_{-0.04}$ & $2.18^{+0.60}_{-0.61}$  \\ 
\hline \hline
\vspace{0.01cm}
\end{tabular}}
\title{Table 1: Constants and parameters} \label{table1}
\end{sidewaystable*}

\section{Bag Constant for NS Candidates with Quark Core for Specified Transition Density}
Gravitational mass of hot NS with quark core and without quark core are different. Presence of quark core in hot NS soften the EOS and therefore reduces maximum mass of NS~\cite{Yazdizadeh2011}. The Brueckner-Bethe-Goldstone formalism is used to establish the EOS of the hadronic matter for the investigation of the structural properties of NSs consist of a quark core at zero~\cite{Burgio2002b} and finite temperatures~\cite{Burgio2007}. Basically, a compact star to be $\beta$ stable when it consists of neutrons and some protons satisfying charge neutrality by leptons. From the $\beta$-stable hadronic matter we can get Quark matter in deconfined state. Hadronic EOS can be taken as $p=\omega_h \rho$ and quark EOS can be given by Eq. (\ref{eq3}). The intersection or crossing point of the hadronic EOS and quark EOS will give the required transition point. If $\rho_t$ is the transition density we will have 
\begin{equation}
\rho_t=\frac{b\omega}{4\pi(\omega_h-\omega)},\label{eq42}
\end{equation}
where quark EOS parameter $\omega$ is larger than hadronic EOS parameter $\omega_h$ which implies that quark EOS is more stiffer than hadronic EOS as $\left(\frac{dp}{d\rho}\right)_q>\left(\frac{dp}{d\rho}\right)_h$. 

The Eq. (\ref{eq42}) turns into the following form after incorporating MIT bag model as
\begin{equation}
B=\frac{\rho_t}{4}(1-3\omega_h).\label{eq43}
\end{equation} 

The value of transition point for the transition from nuclear matter to quark matter is yet not known accurately. In general the range for transition density is 0.5-2 $GeV fm^{-3}$~\cite{Muller1985}. A typical estimate of transition density of about 1 $GeV fm^{-3}$  at zero or nearly zero temperature is confirmed by CERN experiments. According to the experiments hadron-quark transition takes place at $\approx 7\epsilon_{0}\approx 1.1$ $GeV fm^{-3}$ where $\epsilon_{0}\approx$ 156 MeV fm$^{-3}$ is normal nuclear matter density. Transition energy density may be assumed or may be determined to study NS properties. We will consider transition density to be $\approx 7\epsilon_{0}$ and  determine possible values of bag constants for different value of $\omega_h$ as shown in Table 2.

\begin{table*}[htbp!]
\centering
\title{Table 2: Values of bag constants  for NS with quark core} \label{table2}
\vspace{0.5cm}
\begin{tabular}{ccc}
\hline
$\omega_h$ & \multicolumn{2}{c}{Bag constant (MeV fm$^{-3})$}\\
\cline{2-3}
& $m_s=0$  & $m_s=$250~MeV \\
& $\&~\omega=1/3$ & $\&~\omega=0.289$\\
\hline
0.05 & 232.05 & 225.77\\
\hline
0.10 & 191.10 & 178.54\\
\hline
0.20 & 109.20 & 84.07\\
\hline
0.21 & 101.01 & 74.63\\
\hline
0.22 & 92.82 & 65.18\\
\hline
0.23 & 84.63 & 55.73\\
\hline
0.24 & 76.44 & 46.29\\
\hline
0.25 & 68.25 & 36.84\\
\hline
\end{tabular}
\end{table*}	

\section{Results and Conclusions}
Using interpolation technique Rahaman et al.~\cite{Rahaman2014} find a mass function which is best fitted to describe the strange star candidates. This function satisfy all the physical conditions for radial distance from 6.2 km to 12.2 km for a strange star. In the present paper we have solved TOV equation for  isotropic spherically symmetric system by homotopy perturbation method and thereafter invoking general linear EOS we find a mass function which is physical from centre to surface of strange star. No particular form of the time-time component of metric and density function is  assumed. With the help of derived mass function and Einstein field equation we get time-time component of metric and density which are finite and positive at centre. Hence we have developed a general non singular solution which describes strange star. To study properties of strange star we use particularly MIT bag model as a special case of our model. We see that all the properties of strange star physically valid as seen from Figs. \ref{Fig1}-\ref{Fig4}. Also from Fig. \ref{Fig5} we conclude that the stellar structure is stable and satisfies all the energy conditions. We utilized MIT bag model to find possible values bag constants of strange stars for given mass and radius. From Tables 1 and 2 we find that lower limit  $B$=41.58~MeV~fm$^{-3}$ for $EXO~0748-676$ and upper limit $B$=319.31~MeV fm$^{-3}$ for $PSR~1937+21$. Another consequence of the satisfaction of Buchdahl's condition lead to the result that the value of bag constant is always less than $ 0.025/R^2$ for a strange star of given radius $R$. The distribution of observed data of mass and radius of strange stars are shown in the graph of mass-radius relation of strange stars. From Fig. \ref{Fig6} we note that strange stars have the range 41.58~MeV fm$^{-3}<B<$319.31~MeV fm$^{-3}$ and the most stars  likely to possess bag constant in the range 67.84~MeV~fm$^{-3}< B <$141.12~MeV~fm$^{-3}$. The distribution of observed maximum mass of stars in the above range of $B$ is shown in Fig. \ref{Fig6}.

We inspect a linear relationship between central redshift and surface redshift from Table 1 as $Z_c=k_{i}Z_{s}$ where $k_i$ is a constant for individual star. The variation of value of $k_i$ is very small and the mean value of $k_i$ is $k=2.010789 \pm 0.073203$. So, for strange star we can assume $Z_c=kZ_s$.  With the help of this relation mass to radius ratio can be expressed in terms of $k$ and $g_{tt}(0)$ as 
\begin{equation}
\frac{2M}{R}= 1- \frac{k^2}{\left[\frac{1}{\sqrt{g_{tt}(0)}}+(k-1)~\right]^2},\label{eq44}
\end{equation}
which implies that compactness of strange star is controlled by the constant $k$ and central geometric component $g_{tt}$. Since $Z_s<2$ we will have $Z_c<2k$ and $g_{tt}(0)> (2k+1)^{-2}$. We get some agreement with experimental observations. For example in case of RX J1856.5-3754 which is an isolated NS. Prakash et al.~\cite{Prakash2003} proposed that it could be a bare strange star with surface redshift $0.35\pm0.15$. In our study we get $Z_s=0.34^{+0.03}_{-0.04}$ for $RX~J1856.5-3754$. Also in a observation of non-detection of pulsation from the star \cite{Ransom2002} it was shown that the mass to radius ratio should be greater than 0.148 which is in agreement with our result 0.22. The selected compact stars of high $Z_s$, central densities can be regarded as strange star candidates~\cite{Ruderman1972,Glendenning1997,Herzog2011}. The maximum value of surface redshift of a stable stellar structure with isotropic perfect fluid can be derived which is $Z_{smax} \approx 0.62$ \cite{Bondi1964a}. In our model we get the surface redshift for $PSR~1937+21$ to be 0.64 which is maximum value as indicated from the Table 1.  Lindblom~\cite{Lindblom1984} find the limit for surface redshift as $0.184 \leq Z_s \leq 0.854$ for NSs of mass $1.4~M_\odot$ and densities below $3 \times 10^{14}~gcm^{-3}$. In our model we find the range for different strange star candidates as $0.20<Z_s<0.64$.
 
With specific choice of the transition density, we calculated bag constant for NS with quark core for different values of hadronic EOS parameter. From Table 2 it is clear that $B$ has large value for soft hadronic EOS and small value for stiff hadronic EOS. This is in agreement with calculated result of Carinhas~\cite{Carinhas1993} who use quark parameter space for NS with quark core and showed that bag constant posses wide range of values upto 432~MeV fm$^{-3}$.   

A theoretical development is needed to find the definite range of bag constant. As we provide strange star model with general quark EOS, the maximum mass of the star can be determined by taking stiff EOS for which $\omega$ and acoustic speed will have large values.

\section*{Acknowledgments}  SR and FR are thankful to the Inter-University Centre for Astronomy and Astrophysics (IUCAA), Pune, India for providing Visiting Associateship under which a part of this work was carried out. SR is also thankful to the authority of The Institute of Mathematical Sciences, Chennai, India for providing all types of working facility and hospitality under the Associateship scheme. FR is grateful to DST-SERB (EMR/2016/000193), Govt. of India for providing financial support. The work by MK was supported by Ministry of Education and Science of the Russian Federation, MEPhI Academic Excellence Project (contract 02.a03.21.0005, 27.08.2013). A part of this work was completed while AA was visiting IUCAA and the author gratefully acknowledges the warm hospitality and facilities there. We are grateful to Prof. I.G. Dymnikova for careful reading of the manuscript and important comments which have helped us to upgrade the manuscript substantially.

\end{document}